\documentclass[superscriptaddress,showpacs,amsmath,amssymb,twocolumn,nofootinbib]{revtex4-2}
\usepackage{amsmath, amsthm, amssymb, bm, braket}
\usepackage[dvips]{graphicx}
\usepackage{color}

\begin{document}
\title{Analytic continuation in coupling constant applied to two-proton emitters}

\author{Tomohiro Oishi}
\email[E-mail: ]{tomohiro.oishi@ribf.riken.jp}
\affiliation{RIKEN Nishina Center for Accelerator-Based Science, Wako 351-0198, Japan}

\author{Masaaki Kimura}
\email[E-mail: ]{masaaki.kimura@ribf.riken.jp}
\affiliation{RIKEN Nishina Center for Accelerator-Based Science, Wako 351-0198, Japan}

\renewcommand{\figurename}{FIG.}
\renewcommand{\tablename}{TABLE}

\newcommand{\bi}[1]{\ensuremath{\boldsymbol{#1}}}
\newcommand{\unit}[1]{\ensuremath{\mathrm{#1}}}
\newcommand{\oprt}[1]{\ensuremath{\hat{\mathcal{#1}}}}
\newcommand{\abs}[1]{\ensuremath{\left| #1 \right|}}
\newcommand{\slashed}[1] {\not\!{#1}} %--- e.g. \slashed{p} = \gamma^{\mu} p_{\mu}.
\newcommand{\crc}[1] {c^{\dagger}_{#1}}
\newcommand{\anc}[1] {c_{#1}}
\newcommand{\crb}[1] {\alpha^{\dagger}_{#1}}
\newcommand{\anb}[1] {\alpha_{#1}}

\def \beq{\begin{equation}}
\def \eeq{\end{equation}}
\def \beqa{\begin{eqnarray}}
\def \eeqa{\end{eqnarray}}
\def \bir{\bi{r}}
\def \ubir{\bar{\bi{r}}}
\def \bip{\bi{p}}
\def \ubip{\bar{\bi{r}}}
\def \adel{\tilde{l}}%--- orbital angular-momentum number for the small component, g(r).
\def \twop{$2p$}

\begin{abstract}
{\noindent
For quantum meta-stable problems with three particles,
the $^{6}$Be and $^{16}$Ne nuclei as two-proton ($2p$) emitters provide a testing field.
Considering the complexity of many-body meta-stable systems, a tractable solver has been on demand.
We apply the analytic continuation in coupling constant (ACCC) to these nuclei,
and confirm that this method can solve such problems consistently to the time-dependent results as well as to the experimental data.
The sensitivity of the $2p$ energies and widths to the proton-proton interaction is also investigated.
The $^{6}$Be shows a smooth shift from short-life to long-life systems according to the proton-proton interaction.
Same is concluded for the $2^+$ resonance of $^{16}$Ne.
In contrast, for the first and second $0^+$ resonances of $^{16}$Ne, their energies and widths
show the avoid-crossing behaviour, due to the coupling of two configurations in the core-proton subsystem, $^{15}$F.
Since it requires only the bound-state solvers, the ACCC can be a low-cost option to solve three or more-body resonances.
}
\end{abstract}

\maketitle

%%%\section{Introduction} \label{sec:intro}
{\it Introduction.}
Resonance is an essential phenomenon in open-quantum meta-stable systems.
In the non-Hermitian quantum mechanics, such resonant states are defined as the eigenstates with complex eigenenergies.
Analytic continuation in coupling constant (ACCC) has been utilized to solve a variety of resonance problems \cite{89Kuku,1997Tanaka,1999Tanaka,2003Aoyama,2006Funaki,2025Zhang}.
One advantage of ACCC is that one does not need to directly deal with the unbound wave functions with the scattering boundary condition.
Instead, one can extrapolate the bound solutions to the unbound region to evaluate the complex eigenenergy.
%%%In the history of meta-stable solvers \cite{89Kuku}, the two-body problems have been often solved, whereas three or more-body problems remain less investigated.

Two-proton (\twop) radioactive emission is a unique radioactive mode in proton-rich nuclei,
in which the Coulomb barrier exists to realize meta-stable states \cite{2012Pfu_rev,08Blank_01,08Blank_02,2009Gri_rev,2019Qi_rev}.
One simple picture for \twop~emitters is of the daughter nuclei and two protons, where three-body resonant problem should be solved.
In our previous works, time-dependent method was utilized to discuss them \cite{2017Oishi, 2014Oishi}.
However, its cost of computation often becomes expensive, and prevents us from the systematic survey.

Recent development of experiments has provided the accumulated data about \twop~emitters \cite{2012Pfu_rev,08Blank_01,08Blank_02,2009Gri_rev}.
From their energy release, lifetime, and \twop~correlations, a deeper understanding of open-quantum systems is obtained. %%%This process also has a link with the nuclear pairing interactions.
Search for the new \twop~emitters is one of the most active topics, where several experimental facilities are involved.
In correspondence, predictions of \twop~emitters with their energies and lifetimes have been on demand.

In this work, we employ the ACCC method to solve the \twop-emitting resonances.
The three-body model developed in previous works \cite{2017Oishi,2014Oishi} is combined with the ACCC method.
By applying it to the light \twop~emitters, $^{6}$Be and $^{16}$Ne, we examine the validity of ACCC.
Note that, in $^{16}$Ne, there exists a multiplet of \twop-emitting resonances with $J^{\pi}=0^+_1$ \cite{2014Brown}, $0^+_2$ \cite{1997Fohl}, and $2^+$ \cite{2015Marganiec, 2015Brown}.
The sensitivity of \twop-emitting energy and width to the proton-proton pairing interaction is also investigated.

%%%\section{model and formalism}
{\it Model and formalism.}
We employ the three-body model utilized in Refs. \cite{2005HS,07Hagi_01,07Bertulani_76,14Hagi_2n,88Suzuki_COSM,1991BE,1997EBH,2014Lorenzo,2010Oishi,2014Oishi,2017Oishi}.
The physical Hamiltonian reads
\beqa
&& \hat{H}_{3B} = \hat{h}(\bir_1)+\hat{h}(\bir_2) +\hat{H}_{pp}(\bir_1,\bir_2), \\
&& \hat{h}(\bir_k) = -\frac{\hbar^2}{2\mu} \frac{d^2}{dr_k^2} + V_{cp}(r_k),  \\ 
&& \hat{H}_{pp}(\bir_1, \bir_2) = f_{pp} v_{pp}(\bir_1,\bir_2) +\frac{\bip_1 \cdot \bip_2}{m_C}.
\eeqa
The spherical core-proton potential $V_{cp}(r_k)$ ($k=1,2$) is given as
\beqa
&& V_{cp}(r_k) = V_{WS}(r_k) +V_{Coul}(r_k) \nonumber \\
&&= \left[ V_0 + (\bi{l} \cdot \bi{s}) W_{ls} \frac{d}{dr_k} \right] f(r_k)  +V_{Coul}(r_k),
\eeqa
where $f(r_k)= \left( 1 + e^{(r-R_0)/a_0} \right)^{-1}$ (Woods-Saxon profile), and $V_{Coul}(r_k)$ indicates the Coulomb potential.
Parameters of $V_{cp}(r_k)$ for $^{6}$Be and $^{16}$Ne are given in
Refs. \cite{2017Oishi} and \cite{2023Oishi_arXiv}, respectively.

The core-proton Sch\"{o}dinger equation reads
$\hat{h} \ket{\phi_a} = e_a \ket{\phi_a}$, where $e_a$ is the eigenenergy of the single-particle level with the quantum numbers $a=\{ n_a, l_a, j_a, m_a \}$,
which denote the nodal quantum number, orbital angular momentum, coupled angular momentum, and magnetic quantum number, respectively.
The single-particle wave function thus reads
\beq
\phi_a(\bir \sigma) =R_{n_a l_a j_a}(r) \left[ Y_{l_a}(\theta, \phi) \otimes \chi_{\frac{1}{2}} (\sigma) \right]^{(j_a m_a)}.
\eeq
The radial part $R_{n_a l_a j_a}(r)$ is numerically solved with Numerov method.
We employ the single-particle orbits up to $l_{\rm cut}=7$ with the cutoff energy, $E_{\rm cut}=24$ MeV.
The continuum states with $e_a >0$ are discretized within a box of $r_{\rm max}=30$ fm radius.
Note that, since the ACCC procedure requires only the bound solutions, the radial box can be smaller than that for the time-dependent calculations \cite{2017Oishi,2023Oishi_arXiv}.

The \twop~wave function $\Psi_{2p}$ is expanded on the anti-symmetrized basis $\left\{ \tilde{\Phi}_{ab}  \right\}$.
That is
\beq
\Psi_{2p} (\bir_1 \sigma_1 ,\bir_2 \sigma_2) = \sum_{ab} U_{ab} \tilde{\Phi}^{(JM,\pi)}_{ab} (\bir_1 \sigma_1 ,\bir_2 \sigma_2),
\eeq
where
\beq
\tilde{\Phi}^{(JM,\pi)}_{ab} (\bir_1 \sigma_1 ,\bir_2 \sigma_2) = \hat{A} \left[ \phi_a (\bir_1 \sigma_1) \otimes \phi_b (\bir_2 \sigma_2) \right]^{(JM,\pi)}.
\eeq
Here $J=\abs{\bi{j}_a +\bi{j}_b}$ (coupled angular momentum), $M=0,\pm 1,\cdots,\pm J$ (magnetic quantum number), $\pi=(-)^{l_a +l_b}$ (parity), and $\hat{A}$ indicates the anti-symmetrization. %%%Notice that $\ket{\tilde{\Phi}^{(JM,\pi)}_{ab}}$ is the eigenstate of the no-correlation Hamiltonian $\hat{h}(\bir_1)+\hat{h}(\bir_2)$.
The expanding coefficients $U_{ab}$ are determined by diagonalizing the  Hamiltonian matrix including proton-proton term $\hat{H}_{pp}$.

The pairing interaction $v_{pp}$ is tuned to reproduce the experimental \twop~energies.
Its formalism and parameters for $^{6}$Be and $^{16}$Ne are given in Refs. \cite{2017Oishi} and \cite{2023Oishi_arXiv}, respectively.
In the following sections, to investigate how this pairing interaction affects the resonance,
we also employ the factor $f_{pp}$ as
\beq
v_{pp}(\bir_1,\bir_2) \longrightarrow f_{pp} v_{pp}(\bir_1,\bir_2), \label{eq:ppfact}
\eeq
where the experimental point is determined with $f_{pp} =1$.
%%%In this case, the \twop~state becomes a single product of two core-proton states.

To apply the ACCC, we introduce an auxiliary potential multiplied by a factor $\lambda$.
That is
\beq
\hat{H}(\lambda) = \hat{H}_{\rm 3B} +\lambda  V_{ex}(\bir_1 ,\bir_2), \label{eq:XTERN}
\eeq
where $V_{ex}(\bir_1 ,\bir_2) $ is an external, attractive potential.
For three-body systems, there are several options of this ACCC term $\lambda  V_{ex}$.
We check the consistency between different options in the next section.

From Eq. (\ref{eq:XTERN}), the original, physical Hamiltonian coincides $\lambda=0$, where the three-body system is unbound.
For $\lambda>0$, the additional attraction becomes active.
With certain value of $\lambda \geq  \lambda_0$, the system starts to have bound-state solutions as $\hat{H}(\lambda) \ket{\Psi_{2p}} = E(\lambda) \ket{\Psi_{2p}}$, where $E(\lambda) \leq 0$.
This eigenenergy is solved by diagonalizing $\hat{H}(\lambda)$.
Then, that is approximated as $E(\lambda) \cong -w^2(\lambda)$, where
\beq
w(\lambda) = \frac{c_1 x +\cdots +c_N x^N}{d_0 +d_1 x +\cdots +d_N x^N},~~~x=\sqrt{\lambda -\lambda_0}.
\eeq
Namely, we employ Pad\'{e} approximation \cite{89Kuku}. %%%The $\lambda_0$ indicates the bound-unbound border: $E(\lambda=\lambda_0)=0$.
We set $N=5$, and $\left\{ c_1 \cdots c_5, d_0, \cdots d_5 \right\}$ are determined by fitting in the bound region ($\lambda_0 < \lambda$).
Then, we analytically continuate $w(\lambda)$ to the physical point ($\lambda=0$), where
$x=\sqrt{-\lambda_0}$ is purely imaginary.
Thus, the complex eigenenergy can be obtained as
\beq
E(\lambda =0) = -w^2(0) = E_R -i\frac{\Gamma_R}{2},
\eeq
where $\Gamma_R$ is the width of resonance.
Before going to the results, we mention the limit of the evaluation accuracy.
The present Pad\'{e} fitting is feasible for $10^{-4}  \lesssim  \Gamma_R \lesssim 1$ MeV.
The lower limit is due to the accuracy of the three-body solver,
whereas the upper limit is in particular from the Pad\'{e} approximation.

%%%\section{results}
%%%\subsection{$^{6}$Be case}
{\it Application to $^{6}$Be.}
First for benchmark, we focus on the $^{6}$Be nucleus by using the $\alpha+p+p$ three-body model \cite{2014Oishi,2017Oishi}.
Experimental data on its \twop-emitting resonance are obtained as $E_{2p,{\rm Expt.}}=1.372(5)$ MeV and $\Gamma_{2p,{\rm Expt.}} = 0.092(6)$ MeV \cite{88Ajzen,91Ajzen}.

%%%%%%%%%%%%%%%%%%%%%%%%%%%%%%%%%%%%%%%%%%%%%%%%%%%%%%%%%%%%%%%%%%%%%%%%%%%
\begin{table}[b] \begin{center}
\caption{Two-proton resonances of the $^{6}$Be and $^{16}$Ne nuclei.
Ref. \cite{2023Oishi_arXiv} is under review now.
The unit is MeV.
}
\label{table:GID}
\catcode`? = \active \def?{\phantom{0}} %define `?' as ' '(one-blank).
  \begingroup \renewcommand{\arraystretch}{1.2}
  \begin{tabular*}{\hsize} { @{\extracolsep{\fill}} llllll }
\hline \hline
 &$J^{\pi}$ & &$f_{pp}$ &$E_{2p}$  &$\Gamma_{2p}$ \\  \hline
$^{6}$Be  &$0^+$  &Option (i)    &$1.0$  &$1.371$  &$0.061$  \\
 & &Option (ii)    &$1.0$  &$1.374$  &$0.054$  \\
 & &Option (iii)    &$1.0$  &$1.371$  &$0.059$  \\
&&Time-dep. \cite{2017Oishi}  &$1.0$  &$1.371$  &$0.056$  \\
&&Expt. \cite{91Ajzen} & &$1.372(5)$   &$0.092(6)$  \\
\hline
$^{16}$Ne &$0^+_{1}$ &Option (i)          &$1.0$   &$1.419$  &$0.007$  \\
&&Time-dep. \cite{2023Oishi_arXiv}  &$1.0$    &$1.401$     &$0.001$  \\
&&Expt. \cite{2014Brown}               & &$1.466(20)$  &$\leq 0.080$  \\
&$0^+_{2}$ &Option (i)   &$1.0$  &$2.649$  &$0.514$  \\
&&Time-dep. \cite{2023Oishi_arXiv}  &$1.0$ &$2.44?$     &$0.82?$  \\
&&Expt. \cite{1997Fohl}  & &$2.1(2)$  &$-$  \\
&$2^+$ &Option (i)  &$1.0$    &$2.425$  &$0.016$  \\
&      &Option (i)  &$0.7$    &$3.090$  &$0.115$  \\
&&Expt. \cite{2015Brown} & &$3.16(2)$  &$0.175(75)$  \\
\hline \hline
  \end{tabular*}
  \endgroup
  \catcode`? = 12 %initialize `?'.
\end{center} \end{table}
%%%%%%%%%%%%%%%%%%%%%%%%%%%%%%%%%%%%%%%%%%%%%%%%%%%%%%%%%%%%%%%%%%%%%%%%%%%

The parameters of $\hat{H}_{3B}$ are given in Ref. \cite{2017Oishi}.
Note that, for the core-proton subsystem $^{5}$Li, one resonance exists in the $p_{3/2}$ channel of protons ($\pi$):
\beq
\hat{h} \ket{\pi,p_{3/2}} =\left[ E_{1p} -i \frac{\Gamma_{1p}}{2} \right] \ket{\pi,p_{3/2}},
\eeq
where $E_{1p} = 1.96$ MeV and $\Gamma_{1p} \cong 1.5$ MeV \cite{91Ajzen}.
Our original core-proton potential is tuned so as to reproduce these values \cite{2014Oishi, 2017Oishi}.
In addition, the three-body (\twop-emitting) resonance is obtained as
\beq
\hat{H}_{\rm 3B} \ket{\Psi_{2p}} =\left[ E_{2p} -i \frac{\Gamma_{2p}}{2} \right] \ket{\Psi_{2p}},
\eeq
where $E_{2p} = 1.371$ MeV and $\Gamma_{2p} = 0.056$ MeV from the time-dependent calculation \cite{2017Oishi}.
Notice that the \twop-resonance energy and width become remarkably smaller than the $1p$-resonance values.
This is a typical product of the pairing interaction \cite{2017Oishi, 2014Oishi}.

As the option (i) of ACCC calculation, the Hamiltonian with the ACCC parameter $\lambda$ reads
\beq
\hat{H}(\lambda) = \hat{H}_{\rm 3B} +\lambda \left[ V_0 f(r_1)+V_0 f(r_2)\right], \label{eq:XYFISA}
\eeq
where $V_0 f(r)$ is the central term of core-proton Woods-Saxon potential used in $\hat{h}(\bir)$.
In addition, we check the consistency with other options, i.e.,
(ii) by changing the core-proton Coulomb potentials, and
(iii) by changing the proton-proton interaction.
In the option (ii), instead of Eq. (\ref{eq:XYFISA}), we employ the ACCC Hamiltonian as
\beq
\hat{H}(\lambda) = \hat{H}_{\rm 3B} -12\lambda \left[ V_{Coul}(r_1)+V_{Coul}(r_2) \right]. \label{eq:XYFISA_COUL}
\eeq
Here the factor $-12$ is used to have the same scale of fitting to the option (i).
In the option (iii), the ACCC Hamiltonian reads
\beq
\hat{H}(\lambda) = \hat{H}_{\rm 3B} +5\lambda v_{pp}(\bir_1,\bir_2), \label{eq:XYFISA_VPP}
\eeq
where the factor $5$ is used for the same purpose.
Distinguish the physical factor $f_{pp}$ in $\hat{H}_{3B}$ from the ACCC parameter $5\lambda$.

%%%%%%%%%%%%%%%%%%%%%%%%%%%%%%%%%%%%%%%%%%%%%%%%%%%%%%%%%%%%%%%%%%%%%%
\begin{figure}[t] \begin{center}
\includegraphics[width = 0.9\hsize]{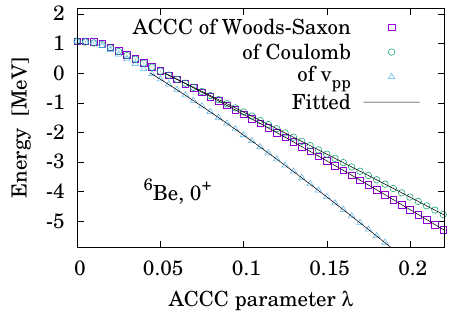}%%%%{Foge_003_06Be_hoppie.pdf}
\caption{Pad\'{e} fitting of the total energy, $E(\lambda)$, of $^{6}$Be with $f_{pp}=1$.
For fitting, calculated results in the bound region with $E< -0.3$ MeV are used.
Results from the ACCC of Woods-Saxon, Coulomb, and proton-proton interactions are compared:
see Eqs. (\ref{eq:XYFISA}), (\ref{eq:XYFISA_COUL}), and (\ref{eq:XYFISA_VPP}), respectively.}
\label{fig:2024_0202}
\end{center} \end{figure}
%%%%%%%%%%%%%%%%%%%%%%%%%%%%%%%%%%%%%%%%%%%%%%%%%%%%%%%%%%%%%%%%%%%%%%

In FIG. \ref{fig:2024_0202}, results of ACCC are presented:
the three-body eigenenergies for $J^{\pi}=0^+$ with variation of external potentials are plotted with symbols.
One can find that, in all the cases, the eigenenergy is a smooth function of $\lambda$.
From Pad\'{e}-approximation fitting, whose result is plotted with line, the parameters of $w(\lambda)$ are determined.
Then we extrapolate this function to the physical point ($\lambda=0$).
In the Woods-Saxon-ACCC case, we obtained $E_{2p}=1.371$ MeV and $\Gamma_{2p} = 0.061$ MeV.
In addition,
the option (ii) by changing $V_{Coul}$ yields $E_{2p}=1.374$ MeV and $\Gamma_{2p} = 0.054$ MeV;
the option (iii) by changing $v_{pp}$ yields $E_{2p}=1.371$ MeV and $\Gamma_{2p} = 0.059$ MeV.
Consequently, the \twop~resonance can be commonly obtained with the three options.
These values are well consistent to the time-dependent results for the same problem \cite{2017Oishi}: see TABLE \ref{table:GID}.
Therefore, the ACCC seems feasible to \twop-emitting resonances.
In the following, the option (i), i.e., Woods-Saxon ACCC by Eq. (\ref{eq:XYFISA}), is used as default.

%%%%%%%%%%%%%%%%%%%%%%%%%%%%%%%%%%%%%%%%%%%%%%%%%%%%%%%%%%%%%%%%%%%%%%
\begin{figure}[t] \begin{center}
\includegraphics[width = 0.9\hsize]{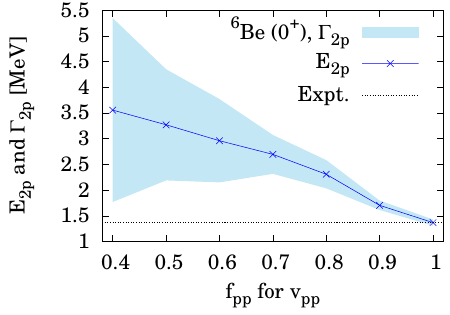}%%%{Foge_001_hogv2.pdf}
\caption{Two-proton $0^{+}$ resonance solved with ACCC for the $^{6}$Be$=^{4}$He$+p+p$.
Here $f_{pp}$ is the proton-proton interaction factor given in Eq. (\ref{eq:ppfact}).
}
\label{fig:2024_0204}
\end{center} \end{figure}
%%%%%%%%%%%%%%%%%%%%%%%%%%%%%%%%%%%%%%%%%%%%%%%%%%%%%%%%%%%%%%%%%%%%%%

Next we confirm the sensitivity of \twop-resonance energy and width to the proton-proton interaction.
It is well known that, with the stronger pairing,
the resonance energy (width) becomes lower (narrower) \cite{2012Maru,2014Oishi,2016Kobayashi}.

In FIG. \ref{fig:2024_0204}, the \twop-resonance energy and width as functions of $f_{pp}$ are displayed.
First at $f_{pp}=0.4$, our ACCC solution gives $E_{2p} = 3.557$ MeV and $\Gamma_{2p} = 1.787$ MeV.
Then for the experimental energy ($f_{pp}=1$),
our ACCC result reads $E_{2p}=1.371$ MeV and $\Gamma_{2p} = 0.061$ MeV.
Therefore, the behaviour expected from the kinetic effect is reproduced \cite{2012Maru,2014Oishi,2016Kobayashi}:
with the stronger attraction with $f_{pp}>0$, the \twop~energy
becomes reduced, and thus, the tunneling probability is also reduced to realize the smaller width (longer lifetime).
Note that, for $f_{pp} < 0.4$, the width goes beyond the evaluation limit of our ACCC.

%%%%%%%%%%%%%%%%%%%%%%%%%%%%%%%%%%%%%%%%%%%%%%%%%%%%%%%%%%%%%%%%%%%%%%%%%%%
\begin{table}[b] \begin{center}
\caption{Core-proton $s_{1/2}$ and $d_{5/2}$ resonances of the $^{15}$F = $^{14}$O$+p$ nucleus.
The unit is MeV.
}
\label{table:GIS}
\catcode`? = \active \def?{\phantom{0}} %define `?' as ' '(one-blank).
  \begingroup \renewcommand{\arraystretch}{1.2}
  \begin{tabular*}{\hsize} { @{\extracolsep{\fill}} llllll }
\hline \hline
  &  &\multicolumn{2}{l}{$s_{1/2}$}  &\multicolumn{2}{l}{$d_{5/2}$}  \\
      &                    &$E_{p}$      &$\Gamma_{p}$     &$E_{p}$      &$\Gamma_{p} $  \\  \hline
Expt. & \cite{2005Guo}     &$1.23(5)$    &$0.50$-$0.84$~~~  &$2.81(2)$  &$0.30(6)$    \\
      & \cite{2004Goldberg} &$1.23$-$1.37$  &$0.7$          &$2.795(45)$  &$0.325(6)$ \\
      &                     &$1.35$-$1.61$  &$-$  \\
      & \cite{2003Peters}   &$1.51(15)$  &$1.2$  &$2.853(45)$  &$0.34$  \\
      & \cite{91Ajzen}      &$1.47(13)$  &$1.0(2)$  &$2.77(10)$  &$0.24(3)$  \\
This work  &  &$1.277$      &$0.604$      &$2.787$  &$0.264$  \\
\hline \hline
  \end{tabular*}
  \endgroup
  \catcode`? = 12 %initialize `?'.
\end{center} \end{table}
%%%%%%%%%%%%%%%%%%%%%%%%%%%%%%%%%%%%%%%%%%%%%%%%%%%%%%%%%%%%%%%%%%%%%%%%%%%

%%%%%%%%%%%%%%%%%%%%%%%%%%%%%%%%%%%%%%%%%%%%%%%%%%%%%%%%%%%%%%%%%%%%%%
\begin{figure}[t] \begin{center}
\includegraphics[width = 0.9\hsize]{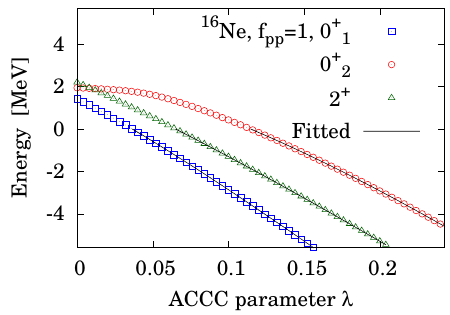}%%%{Foge_002_16Ne_hoppie.pdf}
\caption{Pad\'{e} fitting of the total energy, $E(\lambda)$, of $^{16}$Ne with $f_{pp}=1$.
For fitting, calculated results in the bound region with $E< -0.3$ MeV are used.
For ACCC, the option (i) by Eq. (\ref{eq:XYFISA}) is used: the central part of Woods-Saxon potential is changed.}
\label{fig:2024_1220}
\end{center} \end{figure}
%%%%%%%%%%%%%%%%%%%%%%%%%%%%%%%%%%%%%%%%%%%%%%%%%%%%%%%%%%%%%%%%%%%%%%

%%%\subsection{$^{16}$Ne case}
{\it Application to $^{16}$Ne.}
After a successful benchmark in $^{6}$Be, we apply this ACCC method to another
\twop~emitter $^{16}$Ne, where a complicated energy scheme is measured.
The $^{16}$Ne has several \twop-emitting resonances as summarized in TABLE \ref{table:GID}.
Those include the $0^+_1$ at $1.466(20)$ MeV \cite{2014Brown}, $0^+_2$ at $2.1(2)$ MeV \cite{1997Fohl}, and $2^+$ at $3.16(2)$ MeV \cite{2015Marganiec, 2015Brown}.
We employ the three-body model of $^{16}$Ne $= ^{14}$O$+p+p$ in the following.

In the core-proton subsystem $^{15}$F, two resonances exist in the $d_{5/2}$ and $s_{1/2}$ channels \cite{2005Guo, 2004Goldberg, 2003Peters, 91Ajzen}.
Their experimental data are summarized in TABLE \ref{table:GIS}.
Notice that a large ambiguity remains in the $s_{1/2}$ resonance.
In our model, the core-proton Woods-Saxon and Coulomb potentials are adjusted consistently to Ref. \cite{2005Guo}.
Their parameters are given in Ref. \cite{2023Oishi_arXiv}.
Also, our $v_{pp}$ with $f_{pp}=1$ is the same to Ref. \cite{2023Oishi_arXiv}, and that is adjusted to reproduce the experimental \twop~energy in the $0^+_1$ resonance \cite{2014Brown}.

The ACCC calculation is performed with our option (i) by Eq. (\ref{eq:XYFISA}):
the central part of Woods-Saxon potential is modified.
In FIG. \ref{fig:2024_1220}, the ACCC results with $f_{pp}=1$ are presented.
Consequently, the three \twop~resonances with $0^+_1$, $0^+_2$, and $2^+$ are obtained.
For the Pad\'{e}-approximation fitting, the bound-state energies of $E(\lambda) \leq -0.3$ MeV are utilized.
In TABLE \ref{table:GID}, their energies and widths obtained at $\lambda=0$ are summarized.
For the two $0^{+}$ solutions, their results are roughly consistent to the time-dependent calculations in Ref. \cite{2023Oishi_arXiv}.
We have also confirmed that, with $f_{pp}=1$, the $0^+_1$ resonance is $d_{5/2}$-dominant, whereas $0^+_2$ is $s_{1/2}$-dominant.

For the $0^+_1$ state at the experimental point ($f_{pp}=1$),
we obtain $E_{2p} = 1.419$ MeV with $\Gamma_{2p} = 0.007$ MeV.
The obtained width is smaller than the existing data \cite{2014Brown}.
Note that these data are limited due to the resolution limit.
The exact measurement of keV-order width is still challenging.

%%%%%%%%%%%%%%%%%%%%%%%%%%%%%%%%%%%%%%%%%%%%%%%%%%%%%%%%%%%%%%%%%%%%%%
\begin{figure}[t] \begin{center}
\includegraphics[width = 0.9\hsize]{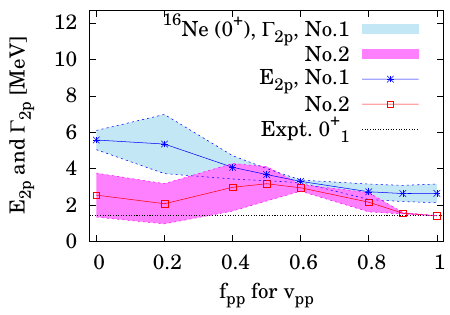}%%%{Foge_002_UD.pdf}
\caption{Two-proton resonances with $J^{\pi} =0^+$ solved with ACCC for the $^{16}$Ne$=^{14}$O$+p+p$.
The $f_{pp}$ is the proton-proton interaction factor in Eq. (\ref{eq:ppfact}).
The label No. 1 (No. 2) indicates the $0^+_1$ ($0^+_2$) resonance.
}
\label{fig:2024_0207_True}
\end{center} \end{figure}
%%%%%%%%%%%%%%%%%%%%%%%%%%%%%%%%%%%%%%%%%%%%%%%%%%%%%%%%%%%%%%%%%%%%%%

%%%%%%%%%%%%%%%%%%%%%%%%%%%%%%%%%%%%%%%%%%%%%%%%%%%%%%%%%%%%%%%%%%%%%%
\begin{figure}[t] \begin{center}
\includegraphics[width = 0.9\hsize]{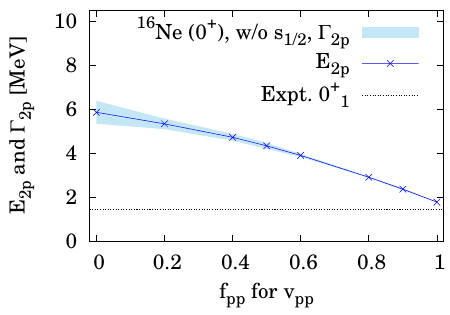}%%%{Foge_005.pdf}
\caption{Same to FIG. \ref{fig:2024_0207_True}, but without $s_{1/2}$ states.}
\label{fig:2024_0503}
\end{center} \end{figure}
%%%%%%%%%%%%%%%%%%%%%%%%%%%%%%%%%%%%%%%%%%%%%%%%%%%%%%%%%%%%%%%%%%%%%%

Next we check the sensitivity to $v_{pp}$.
The results are presented in FIG. \ref{fig:2024_0207_True}.
The $0^+_1$ and $0^+_2$ resonances show an avoid-crossing behaviour.
First in the no-pairing limit ($f_{pp}=0$), we confirmed that the lower $0^+_1$ resonance is $s_{1/2}$-dominant, whereas the $0^+_2$ resonance is $d_{5/2}$-dominant.
Then, for $f_{pp} > 0.8$, the $0^+_1$ resonance turns to become $d_{5/2}$-dominant, whereas the $0^+_2$ one becomes $s_{1/2}$-dominant.
We also mention that, for $0.6 \leq f_{pp} \leq 0.8$, the present ACCC cannot distinguish the two $0^+$ solutions.
For these two resonances, their parameters ($E_{2p} ,\Gamma_{2p}$) show a non-trivial sensitivity to $f_{pp}$,
where the simple elucidation from the kinetic effect \cite{2012Maru,2014Oishi,2016Kobayashi} does not apply.
One possible reason is due to the coupling between $s_{1/2}$ and $d_{5/2}$ channels as well as with non-resonant channels by $v_{pp}$.
For deeper understanding, we aim to report in future studies.

The coexistence of the first and second $0^{+}$ resonances
corresponds to the two core-proton resonances in $^{15}$F.
For this demonstration, we perform the same ACCC calculations but by excluding the $s_{1/2}$ states.
In FIG. \ref{fig:2024_0503}, its result is displayed.
There exists only one resonance, to which the $d_{5/2}$ state is responsible.
In parallel, it does not reach the experimental value without $s_{1/2}$ states:
our result is $E_{2p}=1.794$ MeV, whereas $E_{2p,{\rm Expt.}}=1.466(20)$ MeV \cite{2014Brown}.

%%%%%%%%%%%%%%%%%%%%%%%%%%%%%%%%%%%%%%%%%%%%%%%%%%%%%%%%%%%%%%%%%%%%%%
\begin{figure}[t] \begin{center}
\includegraphics[width = 0.9\hsize]{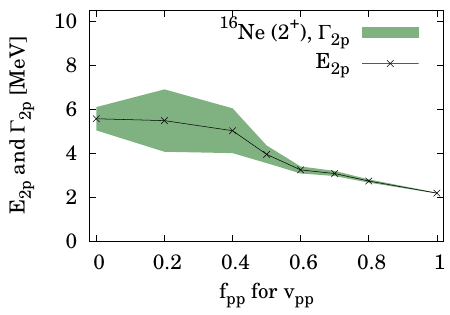}%%%{Foge_008_UD_2P.pdf}
\caption{Same to FIG. \ref{fig:2024_0207_True} but for the $2^+$ state in $^{16}$Ne.}
\label{fig:2024_0207_Two_Plus}
\end{center} \end{figure}
%%%%%%%%%%%%%%%%%%%%%%%%%%%%%%%%%%%%%%%%%%%%%%%%%%%%%%%%%%%%%%%%%%%%%%

%%%\subsection{$2^+$ resonance of $^{16}$Ne}
{\it The $2^{+}$ resonance of $^{16}$Ne.}
This excited resonance of $^{16}$Ne is experimentally observed at $E_{2p} =3.16(2)$ MeV \cite{2015Marganiec, 2015Brown}.
The energy gap between the $2^+$ and $0^+_1$ resonances, i.e., the Thomas-Ehrman shift (TES) \cite{2002Grigorenko, 2015Gri_16Ne, 2015Marganiec}, is connected with the violation of isospin symmetry.
Also, from \twop~correlations measured in this $2^+$ state, so-called
``tethered decaying mechanism'', which cannot be simply attributed to the pure sequential or prompt decay, is suggested \cite{2015Brown}.
For future studies on these topics,
we check the feasibility of ACCC, which enables us to solve the $2^+$ resonance within the similar amount of numerical cost to the $0^+$ case.
%Especially, by considering the isobaric quintet of $(A=16,T=2)$ nuclei,
%TES terms being proportional to $T_z^2$ and $T_z^3$ are suggested \cite{2015Marganiec}.
%Note that such high-order terms of TES should {\it not} be attributed to the Coulomb interaction, whose effect arises only as the $T_z$ term: see Eq. (7) in Ref. \cite{2015Marganiec}.
%Thus, the $2^+$ resonance of $^{16}$Ne can be a good reference quantity to understand the higher-order violation of isospin symmetry.
%In Refs \cite{2002Grigorenko, 2015Gri_16Ne}, this TES is shown as sensitive to the core-proton resonance.

Our results are presented in FIG. \ref{fig:2024_0207_Two_Plus} and TABLE \ref{table:GID}.
For the $J^{\pi}=2^+$ configuration,
the energy and width are obtained as $E_{2p}=2.425$ MeV and $\Gamma_{2p}=0.016$ MeV when $f_{pp}=1$: see TABLE \ref{table:GID}.
Therefore, a finite deviation from the experimental $2^+$-$0^+_1$ gap remains.
We confirmed that this gap is sensitive to the potentials, $V_{cp}$ and $v_{pp}$.
For example, by introducing the $J$-dependent \twop~interaction with $f_{pp}(J=2)=0.7$,
the $2^+$ resonance is reproduced.
The ambiguity in the $s_{1/2}$ resonance of $^{15}$F also affects the $2^+$ energy and width.
Further optimization of potentials may resolve this deviation from experiment.

By changing the interaction strength $f_{pp}$, a smooth shift of $E_{2p}(2^+)$ and $\Gamma_{2p}(2^+)$ is obtained, as shown in FIG. \ref{fig:2024_0207_Two_Plus}.
We checked that the dominant state in this $2^+$ resonance is $d_{5/2}$ at every $f_{pp}$ point, whereas the $s_{1/2}$ state has relatively minor contributions.
Note that, at no-pairing limit ($f_{pp}=0$),
the $E_{2p}(2^+)$ and $\Gamma_{2p}(2^+)$ are approximately identical to the $0^+_2$ case, i.e., they mostly degenerate.
This is because the no-pairing eigenstate is approximately a direct product, $\ket{d_{5/2} \otimes d_{5/2}}$, commonly in the $0^+_2$ and $2^+$ cases.

In the $2^+$ case, only one resonance has been found with the present ACCC method.
On the other hand, in Ref. \cite{2015Marganiec}, the higher $2^+_2$ resonance at $E_{2p} \cong 7.57$ MeV has been reported.
For missing this $2^+_2$ solution of ACCC, its width can be beyond the present limit of evaluation: $\Gamma_{2p} \gtrsim 1$ MeV.

%%%\section{summary} \label{sec:summary}
{\it Summary.}
In this work, the \twop-emitting resonances in the $^{6}$Be and $^{16}$Ne nuclei are investigated.
The ACCC method is developed to solve the resonance energies and widths.
The $^{6}$Be nucleus shows a smooth shift from short-life to long-life systems according to the proton-proton interaction $v_{pp}$.
In contrast, in the $0^+_1$ and $0^+_2$ resonances of $^{16}$Ne, the \twop~energies and widths show complicated behaviours, due to the coupling between different core-proton resonances by $v_{pp}$.
For such a case, we aim to report more careful discussions in future.
We also confirmed the feasibility of ACCC to the excited $2^+$ resonance of $^{16}$Ne.
Although finite deviations from experimental energy remain,
we expect that those are improved by optimizing the potential parameters, independently of the validity of the ACCC.
Consequently, the ACCC method can be used to find the \twop-resonance solutions in various nuclei, including both ground and excited states.

The application of ACCC to other systems is in progress.
Since the ACCC method enables one to find the unbound solutions by solving only the bound states,
its application to other theoretical models may provide a wider survey along the proton-drip line.

%%%\begin{acknowledgments}
{\it Acknowledgments}.
Numerical calculations are supported by the Multi-disciplinary Cooperative Research Program (MCRP) in FY2023 and FY2024 by Center for Computational Sciences, University of Tsukuba (project ID wo23i034), allocating computational resources of supercomputer Wisteria/BDEC-01 (Odyssey) in Information Technology Center, University of Tokyo.
We appreciate the cooperative project of supercomputer Yukawa-21 in Yukawa Institute for Theoretical Physics, Kyoto University.
%%%\end{acknowledgments}


%---
%%\bibliographystyle{apsrev4-2}
%%\bibliography{zb_all}

%%\newpage
%%(To be continued.)
%%\appendix
%%\input{ACCCEW_Append}

\end{document}